\documentclass[aps,prl,showpacs,twocolumn,groupedaddress]{revtex4}  

\usepackage{graphicx}  
\usepackage{dcolumn}   
\usepackage{bm}        
\usepackage{amssymb}   
\input{epsf}

\def\lsim{\mathrel{\rlap{\lower4pt\hbox{\hskip1pt$\sim$}}\raise1pt\hbox{$<$}}}
\def\gsim{\mathrel{\rlap{\lower4pt\hbox{\hskip1pt$\sim$}}\raise1pt\hbox{$>$}}}

\begin{document}


\hspace{5.2in} \mbox{FERMILAB-PUB-07-641-E}
\title{Search for $ZZ$ and $Z\gamma^{*}$ production 
in $p\bar{p}$ collisions at $\sqrt{s}$~=~1.96~TeV 
and limits on anomalous 
$ZZZ$ and $ZZ\gamma^*$ couplings }
%
\author{V.M.~Abazov$^{36}$}
\author{B.~Abbott$^{76}$}
\author{M.~Abolins$^{66}$}
\author{B.S.~Acharya$^{29}$}
\author{M.~Adams$^{52}$}
\author{T.~Adams$^{50}$}
\author{E.~Aguilo$^{6}$}
\author{S.H.~Ahn$^{31}$}
\author{M.~Ahsan$^{60}$}
\author{G.D.~Alexeev$^{36}$}
% ...

\affiliation{\vspace{0.1 in}(The D\O\ Collaboration)\vspace{0.1 in}}
% ...
\affiliation{$^{6}$University of Alberta, Edmonton, Alberta, Canada,
                Simon Fraser University, Burnaby, British Columbia, Canada,
                York University, Toronto, Ontario, Canada, and
                McGill University, Montreal, Quebec, Canada}
% ...
\affiliation{$^{29}$Tata Institute of Fundamental Research, Mumbai, India}
% ...
\affiliation{$^{31}$Korea Detector Laboratory, Korea University, Seoul, Korea}
% ...
\affiliation{$^{36}$Joint Institute for Nuclear Research, Dubna, Russia}
% ...
\affiliation{$^{50}$Florida State University, Tallahassee, Florida 32306, USA}
% ...
\affiliation{$^{52}$University of Illinois at Chicago,
                Chicago, Illinois 60607, USA}
% ...
\affiliation{$^{60}$Kansas State University, Manhattan, Kansas 66506, USA}
% ...
\affiliation{$^{66}$Michigan State University,
                East Lansing, Michigan 48824, USA}
% ...
\affiliation{$^{76}$University of Oklahoma, Norman, Oklahoma 73019, USA}
% ...

\date{December 3, 2007}

\begin{abstract}
We present a study of  $\mu\mu\mu\mu$, $eeee$, and $\mu\mu ee$ events
using 1~fb$^{-1}$ of data collected with the D0~detector at the
Fermilab Tevatron $p\bar{p}$ Collider at $\sqrt{s}$ = 1.96 TeV.
Requiring the lepton pair masses to be greater than 30~GeV, we
observe one event, consistent with the expected background of $0.13\pm 0.03$
events and with the predicted standard model $ZZ$ and $Z\gamma^{*}$ 
production of $1.71\pm0.15$ events.  We
set an upper limit on the $ZZ$ and $Z\gamma^{*}$ cross section of 4.4~pb at the
95\%~C.L.  We also derive limits
on anomalous neutral trilinear $ZZZ$ and $ZZ\gamma^*$ gauge couplings.
The one-parameter $95\%$ C.L. coupling
limits with a form factor scale $\Lambda$
 = 1.2~TeV are $-0.28 <f_{40}^Z<0.28$, 
$-0.31<f_{50}^Z<0.29$, 
$-0.26<f_{40}^\gamma<0.26$,
and $-0.30<f_{50}^\gamma<0.28$.
\end{abstract}

\maketitle

The standard model (SM) makes precise predictions for the couplings
between gauge bosons based on the non-Abelian symmetries of the model.
These predictions can be tested by studying di-boson production ($WW$,
$WZ$, $ZZ$, $Z\gamma$, and $W\gamma$) at particle colliders.
$ZZ$ production is predicted to have the smallest cross section among 
the di-boson processes.  Because the decay channels with the smallest expected
background also have the smallest branching fractions, it has not 
been observed at a hadron  collider.  Nevertheless, the final states with 
small backgrounds may provide the best opportunity to observe the
effects of physics beyond the SM.
In addition to the production of new particles that could decay into
$ZZ$, various extensions of the SM predict large anomalous
values of the trilinear couplings~\cite{zz-theory} $ZZZ$ and $ZZ\gamma^{*}$
that would result in higher cross sections than the SM prediction. 
The direct $ZZ\gamma^*$ and $ZZZ$ couplings are zero in the SM.
Consequently, an observation of an enhancement of the cross section 
would indicate physics beyond the SM.

Previous studies of $ZZ$ production 
were made at the LEP electron-positron collider.
The combined LEP results are available in 
Ref.~\cite{lep_combined}. All results are consistent with SM predictions. 
The CDF Collaboration searched for $ZZ$ and $WZ$ production in
$p\bar{p}$ collisions at center-of-mass energy
$\sqrt{s}=1.96$ TeV with the result that
$\sigma(ZZ)+\sigma(WZ)< 15.2$ pb at 95\% C.L.~\cite{CDFzzwz}
The predicted $ZZ$ production cross section in the SM is
$1.6\pm 0.1$ pb at the Tevatron Collider 
energy~\cite{zz-theory,campbellellis}.

In this Letter, we present a search for $ZZ$ and $Z\gamma^{*}$ production 
and a search for anomalous trilinear $ZZ\gamma^{*}$ 
and $ZZZ$ couplings at the Fermilab Tevatron Collider with the
D0 detector. 
We follow the framework of Ref.~\cite{zz-theory}, where general $ZZV$ 
($V$ = $Z$, $\gamma$) gauge and Lorentz-invariant interactions are
considered.  Such $ZZV$ couplings can be parameterized by
two CP-violating ($f_4^V$) and two CP-conserving ($f_5^V$)
complex parameters. Additional anomalous couplings~\cite{glr}  can 
contribute when the $Z$ bosons are off-shell. However, these couplings are 
highly suppressed near the $Z$ boson resonance.
We note that the $ZZV$ couplings are 
distinct~\cite{zz-theory} from the $Z\gamma V$ couplings probed 
in $Z\gamma$ production in $e^+ e^-$ and hadronic collisions.
Partial-wave unitarity is 
ensured by using a form-factor parameterization that causes the
coupling to vanish at high parton center-of-mass energy $\sqrt{\hat s}$: 
$f_i^V = f_{i0}^{V}/(1+\hat s/\Lambda^{2})^{n}$.
Here, $\Lambda$ is a form-factor scale, $f^{V}_{i0}$ are the
low-energy approximations of the couplings, and $n$ is the form-factor
power. In accordance with Ref.~\cite{zz-theory}, we set $n = 3$ for 
all cases. The form-factor scale $\Lambda$ is selected so that limits 
are within the values provided by unitarity at Tevatron Collider energies. 

We search for $ZZ$ and $Z\gamma^{*}$ 
production with a final state signature that
consists of four leptons:  two pairs of either electrons or muons. 
The electron and muon pairs can be produced either by
the decay of an on-shell $Z$ boson or via a virtual $Z$ boson or 
photon.  We studied three final states: four muon ($\mu\mu\mu\mu$),
four electron ($eeee$), and two muons and two electrons  ($\mu\mu ee$). 

Data used in this analysis~\cite{jarvis} were collected with the D0~detector
at the Fermilab Tevatron $p\bar{p}$ collider at 
$\sqrt{s}=1.96$ TeV between October 2002 and February 2006.
The integrated luminosities~\cite{d0lumi} of the three channels 
are approximately 1 fb$^{-1}$ each and are shown in 
Table~\ref{tab:channels}.

The D0~detector~\cite{run2det} is a multi-purpose detector designed 
to operate at high luminosity. The main components
of the detector are an inner tracker, a liquid-argon/uranium calorimeter,
and a muon system. The inner tracker consists of a 
silicon microstrip tracker (SMT) and a central fiber tracker (CFT) operating
in a 2 Tesla solenoidal magnetic field. 
The SMT has coverage to pseudorapidity~\cite{pseudo}
$|\eta| \approx$~3.0,
while the CFT has coverage to $|\eta| \approx$~1.8.
The calorimeter system is outside the solenoid and 
has three cryostats, one for each of the two
endcap calorimeters (EC) and one for the central calorimeter (CC).
 The CC covers
the region $|\eta| < 1.1$, while endcap calorimeters extend the coverage
to $|\eta| \approx$ 4.
The calorimeters are segmented along the shower direction with 
four layers forming the electromagnetic section (EM) and an additional
three to five layers forming the hadronic section. 
This allows for electron-pion 
separation based on longitudinal and transverse shower shape.
 The muon system is outside of the calorimeters and
consists of 1.8 Tesla iron toroid magnets with tracking chambers and 
scintillator counters mounted both inside and outside the toroid iron. 
The muon system extends to $|\eta| \approx$2.

The D0~detector utilizes a three-level (L1, L2, and L3) trigger system.
Electrons candidates are required to have deposited energy in the 
EM section of the calorimeter at L1 and L2, and then to satisfy selection
criteria on shower shape and the fraction
of energy in the EM calorimeter at L3. Muon candidates
are required to have hits in the muon scintillation
counters and a match with a track in 
the L1 track system. In a fraction of the data, muons 
are also required to have hits in the chambers of the
muon system. At L2, muon candidates are required to have track segments 
in the muon tracking detectors,
and a minimum transverse momentum ($p_T$) is required. At L3, some muons 
are also required to have
a match with a track from the inner tracker. 

Combinations of single-electron 
and di-electron, as well as single-muon triggers are used in this analysis.
These triggers have varying $p_T$ and quality requirements
on the leptons. The trigger efficiency for events with four high-$p_T$
leptons that satisfy all other selection requirements is approximately 99\%.
In order to increase the collection efficiency, we do not require pairs
of leptons to have opposite electric charge.

For the $\mu\mu\mu\mu$ channel, each muon is required to satisfy 
quality criteria
based on scintillator and wire chamber information from the muon
system and to have a matching track in the inner tracker.  The track
matched to the muon must have $p_T > 15$~GeV.  
Additionally, muons that 
are only identified in the muon detector layers before the toroid are
required to have less than $2.5$~GeV of total transverse energy
deposited in the calorimeter in an annulus
between $R=0.1$ and $R=0.4$ centered around the
track matched to the muon~\cite{cal_iso}. 
Finally, the distance in the transverse plane 
of closest approach to the beamspot 
for the track matched to the muon must be less than 0.02~cm
for tracks with SMT hits and less than 0.2~cm for tracks without SMT
hits. This reduces the background from muons that do not originate from the
primary vertex such as those from $b$ decays and cosmic rays. The
maximum distance between the muon track vertices for all muon pairs
along the beam axis is required to be less than 3~cm. This reduces
backgrounds from pairs of cosmic ray muons and from beam halo.
Since the charge of the muons is not considered, 
three possible
$Z/\gamma^{*}$ combinations can be formed for each $\mu\mu\mu\mu$
event. Selected events are required to have the invariant masses
of both muon pairs above 30~GeV for at least one of the 
combinations.

For the $eeee$ channel, each electron is required to have 
transverse energy $(E_{T})$ greater than $15$~GeV,
and to have $|\eta| < 1.1$ or $ 1.5 < |\eta| < 3.2$, and to be
isolated from other energy clusters. Electrons are also
required to satisfy identification criteria based on multivariate
discriminators derived from calorimeter shower shape and tracking
variables. Since the calorimeter fiducial region covers regions where
there is little or no tracking coverage, only three of the four
electrons are required to have an associated inner track.
Electrons without a track match are required to satisfy more
stringent shower
shape requirements. Similar to the $\mu\mu\mu\mu$ channel,
events are required to have the invariant masses
of both electron pairs to be above 30~GeV for at least one of the
three possible combinations.
 
For the $\mu\mu ee$ channel, muons are required to satisfy the same
single muon selection criteria as in the $\mu\mu\mu\mu$ channel and
electrons the same transverse energy
and $\eta$ criteria as in the $eeee$ channel. Both electrons are
required to satisfy the multivariate discriminator based on shower shape
and parameters of the matching inner track. In addition, electrons 
and muons are
required to be separated by $R > 0.2$ to reduce backgrounds from
muons that have radiated photons spatially coincident with a track.
The invariant masses of the muon pair and the electron pair are
required to be greater than 30~GeV.

We determine the total acceptance using a combination
of information from MC and the data. 
We determine single electron and muon identification
efficiencies directly from data~\cite{tag_and_probe}, which  
contains high $p_T$ electrons
and muons from more than 100,000 single $Z$ decays in each channel.
Efficiencies are parameterized as functions of the relevant
variables for the tracking, calorimeter and muon systems such as
the position of the interaction along the beam direction and the
$\eta$ of the electron or muon. The acceptance for all channels is then
determined using $ZZ$ and $Z\gamma^{*}$ events generated 
with {\sc pythia}~\cite{Pythia}
and a parameterized simulation of the detector.
The momenta of the muons and electrons are fluctuated in the 
Monte Carlo (MC) using
angular and energy resolutions determined from the data,
and the measured single electron and muon identification efficiencies
are taken into account in calculating the acceptance.
The number of observed events, the acceptance, and the expected signal
calculated assuming the cross section is 1.6 pb 
are listed in  Table~\ref{tab:channels} for each channel. 
Systematic uncertainties in acceptance due to momentum and energy scale
      calibrations, angular resolution, lepton identification variation
      with luminosity as well as other effects were included.
Taking into account correlations between these uncertainties,
a total of $1.71\pm0.15$ $ZZ$ and $Z\gamma^*$ events is expected.
\begin{table*}
\begin{tabular}{|c|c|c|c|c|} \hline \hline
Decay Channel & Luminosity (pb$^{-1}$)
                         & Acceptance & Expected Signal  
                                                       & Observed Events 
                                                           \\ \hline
$\mu\mu\mu\mu$& $944\pm 58$
                         & $0.27\pm 0.02$
                                      & $0.46\pm 0.05$ & 0  \\ \hline
eeee          & $1070\pm 65$
                         & $0.23\pm 0.01$
                                      & $0.44\pm 0.03$ & 0  \\ \hline
$\mu\mu$ee    & $1020\pm 62$
                         & $0.22\pm 0.02$
                                      & $0.81\pm 0.09$ & 1  \\ \hline \hline
\end{tabular}
\caption{The integrated luminosity, acceptance (including lepton
identification efficiencies),  expected number of signal
candidates, and the number of observed events in the three decay channels
for the $ZZ$ and $Z\gamma^*$ cross section analysis.  
A total of $1.71\pm 0.15$ $ZZ$ and $Z\gamma^*$ events is expected. }
\label{tab:channels}
\end{table*}

The main background sources are ``$Z$ + multi-jet'' events, 
events where a top-antitop ($t\bar{t}$) quark pair is
produced, and ``combinatoric events'', which are four-lepton events 
that survive because mis-paired leptons
cause them to satisfy the dilepton invariant mass selection criteria 
when they would not have otherwise.

For a ``$Z$ + multi-jet'' event to be a $ZZ$ candidate,
the jets are either mis-identified as electrons or muons or
contain real electrons or muons from in-flight decays of pions,
kaons, or heavy quarks.  Though $Z$ + jets  (and $\gamma^*$ + jets) 
events are the primary source
of this background, it also includes those events where the $Z$ boson
or $\gamma^*$  is reconstructed from one or more ``fake'' leptons.
The ``$Z$ + multi-jet'' event background is measured by first 
determining the probability for a jet in data to produce an 
electron or muon that satisfies the identification criteria. 
The probability for a jet to mimic a muon was measured in two-jet
events selected via jet triggers. 
Muons that satisfy the selection criteria, have $p_T > 15$ GeV, and are 
found within $R<0.2$ of the lower-$E_T$ jet are considered ``fake''.
The probability for a jet to mimic a muon is parameterized in
terms of the tag-jet $E_T$. 
It is $10^{-4}$ for jets with $E_T \approx 15$ 
GeV and increases approximately linearly with $E_T$ to $10^{-2}$
for jets with $E_T \approx 150$ GeV. 
The probability for a jet to mimic an electron was measured 
using a procedure similar to that described for muons. 
It is parameterized in terms of $\eta$ and $E_T$ to account for 
differences in the EC and CC and is typically  
$10^{-4}$ ($10^{-2}$) per jet for those electrons 
with (without) a track match.
A systematic uncertainty of $30\%$ is assigned to account for
variations in the probabilities as a result of changes of the lepton 
identification criteria that  were performed as cross checks. 
The probabilities for jets to be misidentified as muons are then 
applied to the jets in our $\mu \mu$ + jets data to determine the background
to $\mu\mu\mu\mu$. Similarly appropriate probabilities are applied to
$\ell\ell\ell$ + jet(s) events to determine the background 
to $eeee$ and $\mu\mu ee$. There, because we started from a three lepton sample,
we correctly account for events with two real leptons, a photon 
misidentified as an electron, and a jet misidentified as either 
an electron or muon.

The background from $t\bar{t}$ events is determined from {\sc pythia} 
MC using a detailed 
{\sc geant}-based~\cite{geant} detector simulation program
and the same reconstruction program that was used for the data. 

The combinatoric background occurs only in the four-electron
and four-muon decay channels. While it could be reduced by 
$\sim 1/3$ by requiring
leptons which form $Z$ bosons to have opposite signs, that 
would have resulted in a loss of signal efficiency for the 
high-$p_T$ leptons that could result from anomalous couplings.
This background was estimated using MC simulation.   
It is $0.016\pm0.003$ ($0.015\pm0.003$) events
in the $\mu\mu\mu\mu$ (eeee) channel; the uncertainty in the 
background comes from uncertainty in the lepton $p_T$ resolution.

Table~\ref{Table:backgrounds} displays the contributions of
all non-negligible backgrounds and a summary for each decay 
channel. The total expected number of candidates from background
sources is $0.13\pm0.03$ events. 
\begin{table}
\begin{tabular}{|c|c|c|c|} \hline \hline
Background                & $\mu\mu\mu\mu$  & $eeee$       &     $\mu\mu ee$  
                                                            \\ \hline \hline
$Z$ + multi-jet            & $0.004\pm 0.001$& $0.065\pm 0.021$
                                                           & $0.007\pm 0.002$
                                                           \\ \hline
$t\bar{t}$                & $0.010\pm 0.005$& -
                                                           & $0.006\pm 0.003$    \\ \hline
Combinatoric 
                           & $0.016\pm 0.003$& $0.015\pm 0.003$ 
                                                           & -  
                                                            \\ \hline
Beam Halo                  & $0.003\pm 0.001$&-            &- 
                                                            \\ \hline \hline
Total                      & $0.033\pm 0.006$&$0.080\pm 0.021$
                                                           & $0.013\pm 0.004$ 
                                                            \\ \hline \hline  
\end{tabular}
\caption{Contributions of non-negligible backgrounds to the expected number
of candidates in the three decay channels for the cross section analysis.
The total expected background is  $0.13\pm 0.03$ events. }
\label{Table:backgrounds}
\end{table}

One event is observed in data, consistent with the SM  prediction
of $1.71\pm 0.15$ events plus background of $0.13\pm0.03$ events.
We set an upper limit of $4.4$~pb at the 95\%~C.L. on the cross section for 
$p\bar{p}\rightarrow ZZ+X$ and $Z\gamma^*+X$, where dilepton pair
masses are greater than $30$ GeV.

Because we do not observe an excess of events,
we set limits on anomalous trilinear couplings by comparing the number
of observed candidates with the predicted background and the 
expected sum of $ZZ$ events from a MC program~\cite{zz-theory} 
using anomalous $ZZ\gamma^*$ and $ZZZ$ couplings. 
We produced grids of MC samples by varying two of the anomalous couplings 
at a time. There are typically 5000 events generated at each grid point. 
These events are processed through the same detector and reconstruction 
simulation as was used in the cross section analysis.
Because on-shell $Z$ bosons dominate the contributions of the
$ZZV$ couplings in the MC and because off-shell $Z$ bosons
are enhanced by couplings~\cite{glr} not implemented in the MC,
we required the dilepton invariant 
mass to be $> 50$ $(70)$ GeV for dimuons (dielectrons). 
 This selection criterion is set by the resolution 
of the dilepton mass measurement and it removed the single 
$\mu\mu ee$ candidate event. 

\begin{figure}[htb]
\hbox{
\epsfxsize = 1.7in
\epsffile{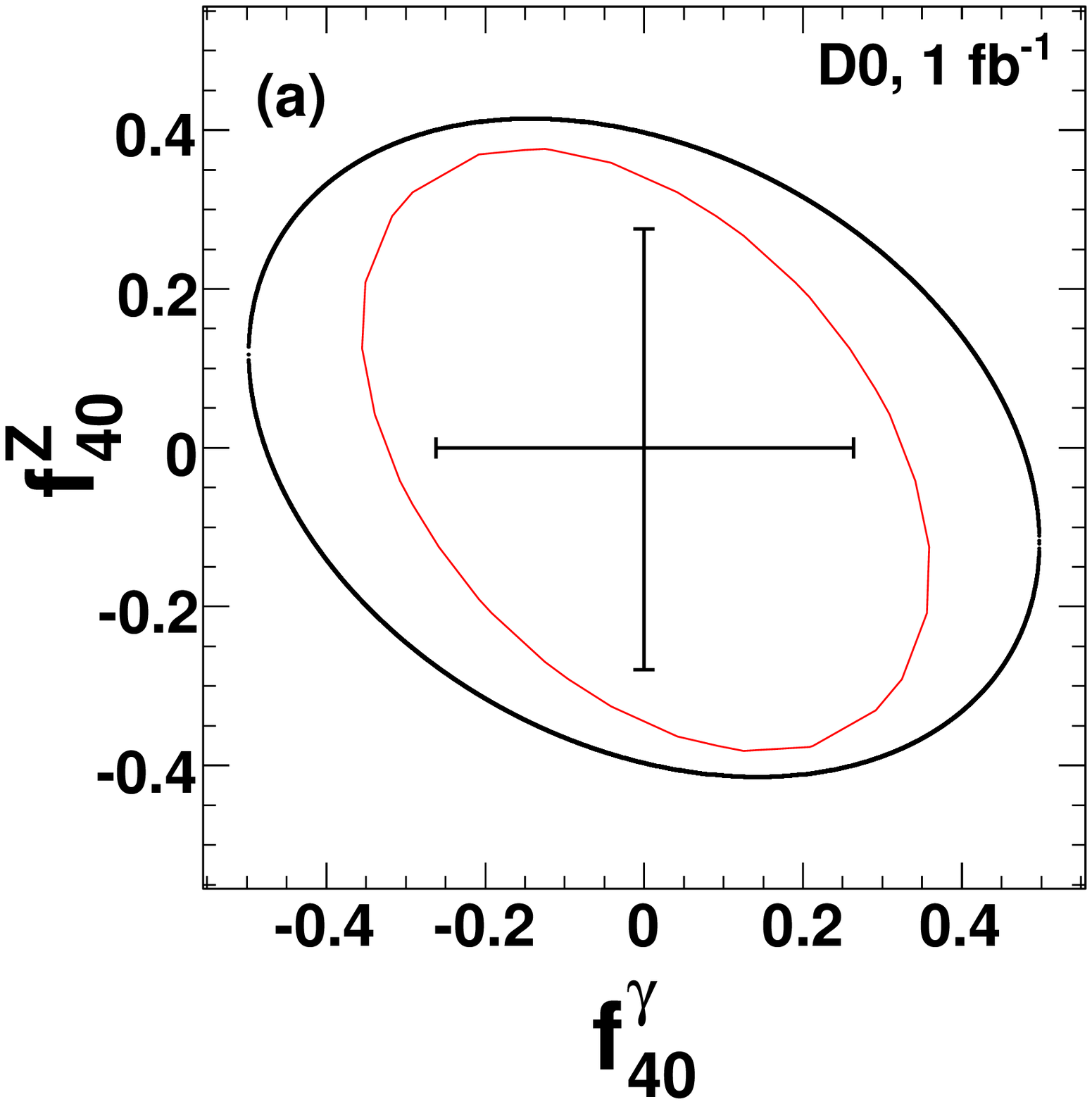}
\epsfxsize = 1.7in
\epsffile{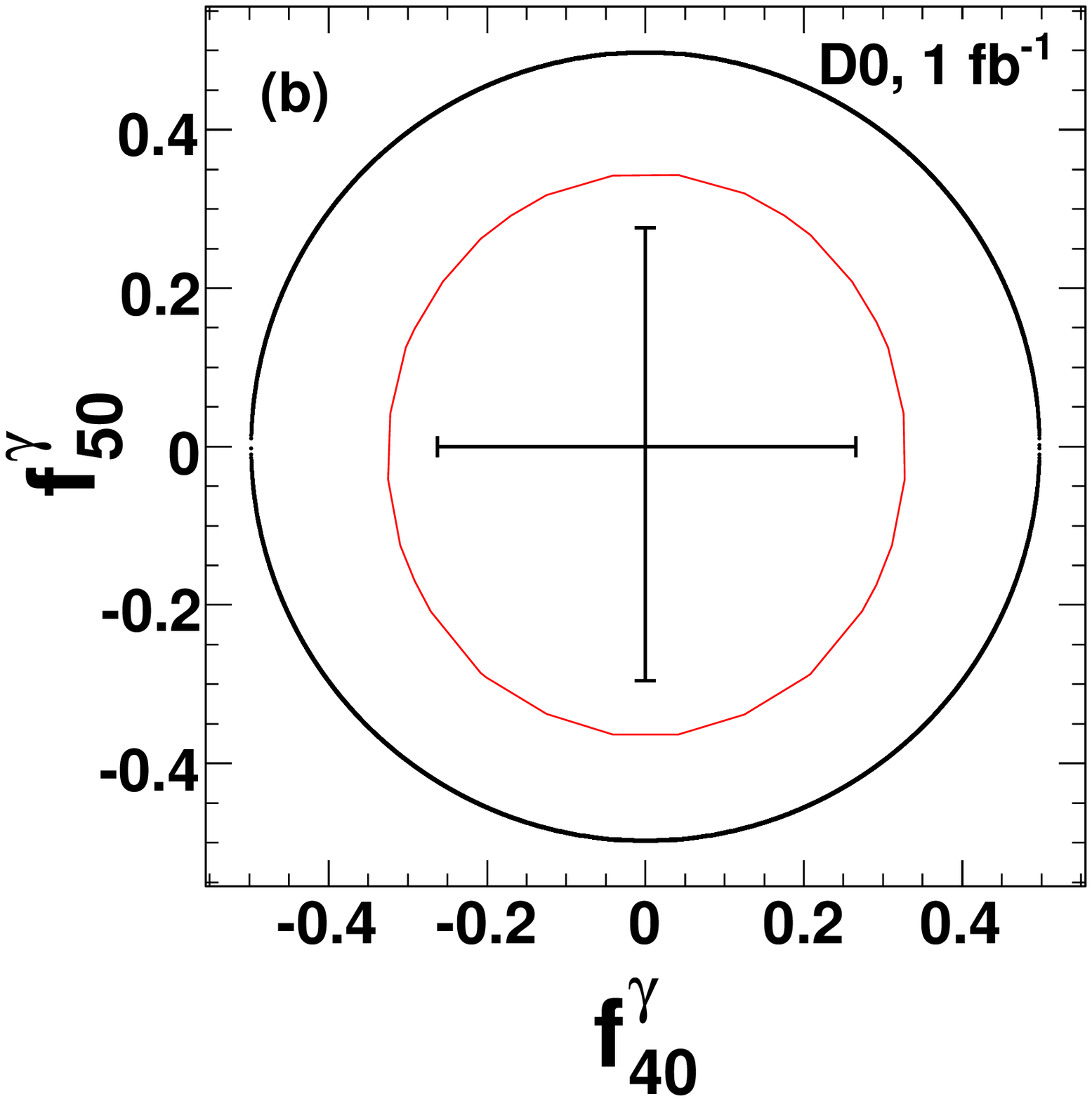}  }
\hbox{
\epsfxsize = 1.7in
\epsffile{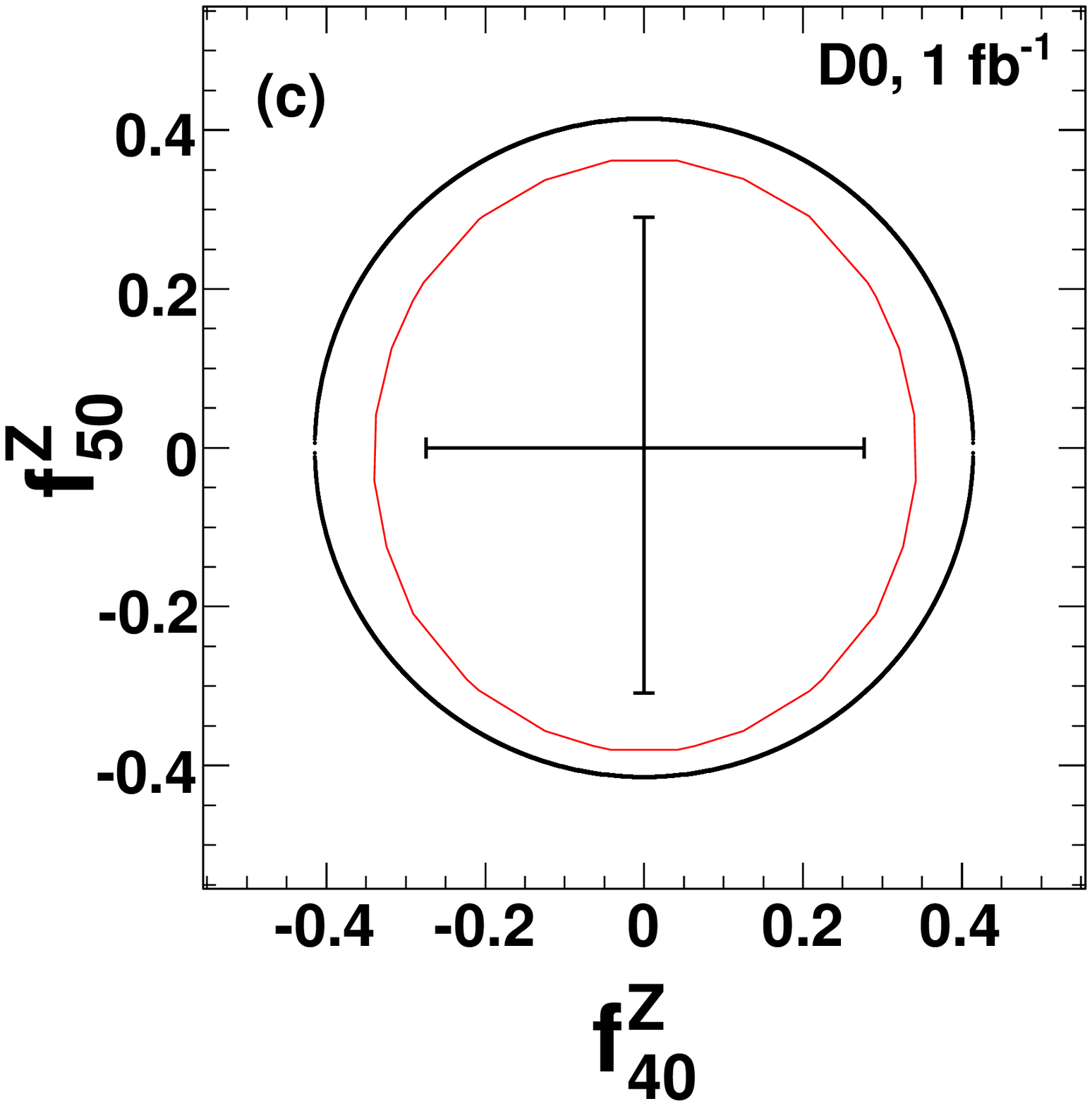}
\epsfxsize = 1.7in
\epsffile{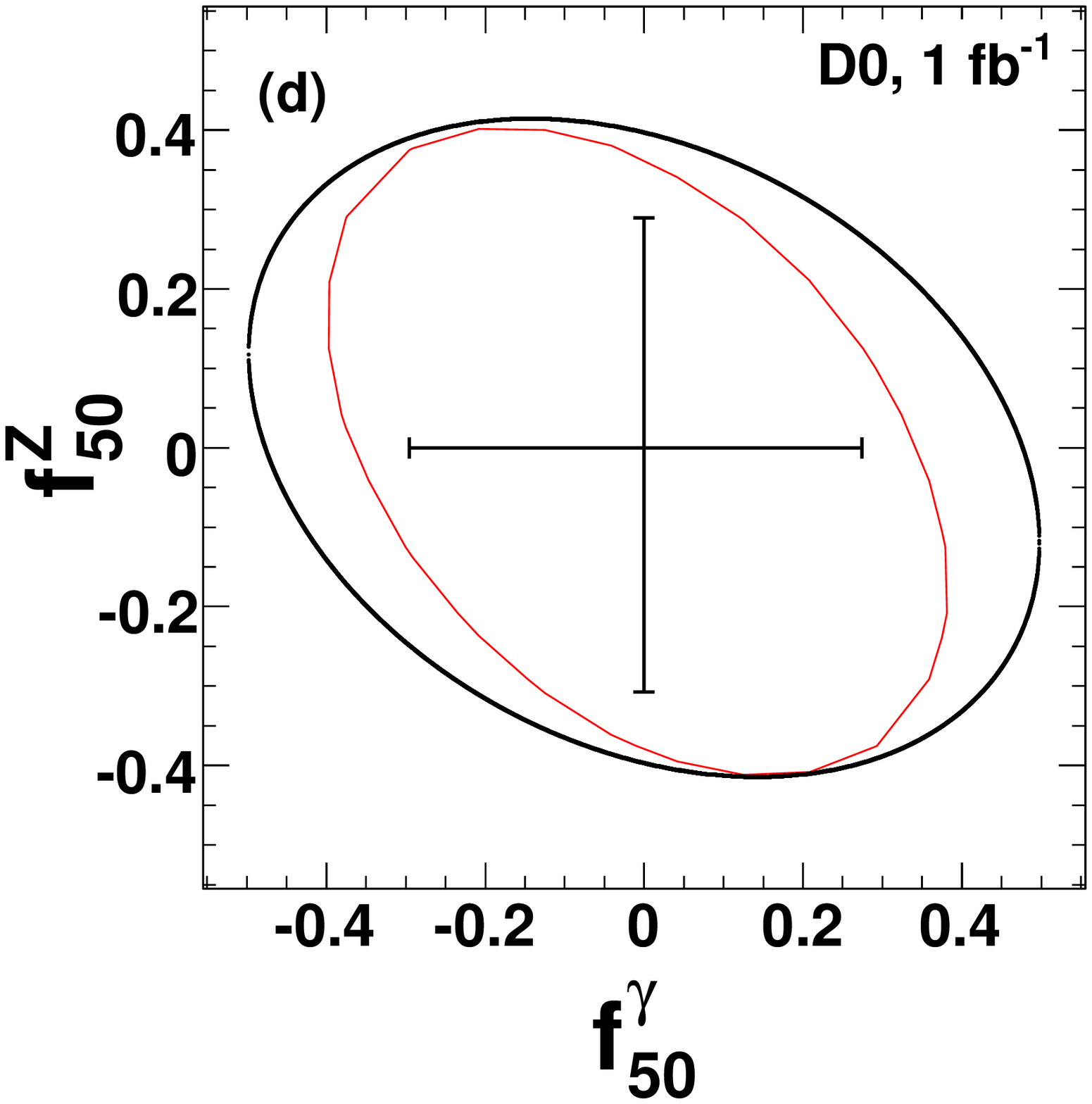} }
\caption{Limits on anomalous couplings for $\Lambda = 1.2$ TeV: 
(a) $f_{40}^{\gamma}$ vs. $f_{40}^Z$,
(b) $f_{40}^{\gamma}$ vs. $f_{50}^{\gamma}$,
(c) $f_{40}^Z$ vs. $f_{50}^Z$, and
(d) $f_{50}^{\gamma}$ vs. $f_{50}^Z$,
 assuming in each case that the other two couplings are zero.
 The inner and outer curves are the
 $95 \%$ C.L. two-degree of freedom exclusion contour
 and the constraint from the unitarity condition, respectively.
 The inner crosshairs are the  $95 \%$ C.L. one-degree of freedom 
exclusion limits. }
\label{fig:combined}
\end{figure}

The number of events expected for each choice of anomalous 
couplings was used to form a likelihood~\cite{jarvis} for that point. 
Poisson probabilities were used for the expected signal plus background.
These are convoluted with Gaussian uncertainties on the acceptance,
background, and luminosity. One-dimensional (1D) and two-dimensional (2D) 
limits on anomalous couplings are formed by finding the coupling values 
with likelihood of 1.92 and 3.00 units greater than the minimum. 
Limits are determined using $\Lambda = 1.2$~TeV.
The 95\% C.L. 1D limits are 
$-0.28 < f_{40}^Z < 0.28$, $-0.26 < f_{40}^{\gamma} < 0.26$,
$-0.31 < f_{50}^Z < 0.29$, and $-0.30 < f_{50}^{\gamma} < 0.28$.
The 95\% C.L. 2D contours 
$f_{40}^{\gamma}$ vs. $f_{40}^Z$, 
$f_{40}^{\gamma}$ vs. $f_{50}^{\gamma}$,
$f_{40}^Z$ vs. $f_{50}^Z$, and 
$f_{50}^{\gamma}$ vs. $f_{50}^Z$
are shown in Figure~\ref{fig:combined}. 
In the four 2D cases the other two couplings are assumed to be zero.
%
%

In summary, we report results from a search for 
$ZZ$ and $Z\gamma^*$ production using  the 
 $eeee$, $\mu\mu\mu\mu$ and $\mu\mu ee$ decay signatures. 
We analyzed 1~fb$^{-1}$ of data collected with the D0~detector at the
Fermilab Tevatron $p\bar{p}$ Collider at $\sqrt{s}$ = 1.96 TeV.
Requiring each lepton to have transverse momentum greater than
15~GeV, and the dilepton pair masses to be greater than 30~GeV, we
observe one event with an expected SM  background of $0.13\pm 0.03$
events.  The one observed event is consistent both with background
and with predicted 
SM  $ZZ$ and $Z\gamma^{*}$ production of $1.71\pm0.15$ events.  
We set an upper limit of $4.4$~pb at the 95\%~C.L. on the cross section for 
$p\bar{p}\rightarrow ZZ+X$ and $Z\gamma^*+X$, where dilepton pair
masses are greater than $30$ GeV. 
This is the most restrictive cross section limit for 
$ZZ$ production at the Tevatron.  We set limits
on anomalous neutral trilinear $ZZZ$ and $ZZ\gamma^*$ gauge couplings.
These represent the first bounds on these anomalous couplings
from the Tevatron. Limits on $f_{40}^Z$, $f_{50}^Z$, and
$f_{50}^{\gamma}$ are more restrictive than those
of the combined LEP experiments~\cite{lep_combined}.

%
We thank the staffs at Fermilab and collaborating institutions, 
and acknowledge support from the 
DOE and NSF (USA);
CEA and CNRS/IN2P3 (France);
FASI, Rosatom and RFBR (Russia);
CAPES, CNPq, FAPERJ, FAPESP and FUNDUNESP (Brazil);
DAE and DST (India);
Colciencias (Colombia);
CONACyT (Mexico);
KRF and KOSEF (Korea);
CONICET and UBACyT (Argentina);
FOM (The Netherlands);
Science and Technology Facilities Council (United Kingdom);
MSMT and GACR (Czech Republic);
CRC Program, CFI, NSERC and WestGrid Project (Canada);
BMBF and DFG (Germany);
SFI (Ireland);
The Swedish Research Council (Sweden);
CAS and CNSF (China);
Alexander von Humboldt Foundation;
and the Marie Curie Program.
%


\begin{thebibliography}{99}
\bibitem{pseudo}
     The D0  coordinate system is cylindrical with the 
     $z$-axis along the proton beamline and the polar and azimuthal
     angles denoted as $\theta$ and $\phi$ respectively. The 
     pseudorapidity is defined as $\eta=-\ln{\tan(\theta /2)}$.
\bibitem{cal_iso} The variable $R$ between two objects
$i$ and $j$ is defined as $R=\sqrt{(\eta_{i}-\eta_{j})^{2} + 
(\phi_{i}-\phi_{j})^{2}}.$
\end{thebibliography}
\end{document}